\documentclass[aps,prd,preprint,groupedaddress]{revtex4}
\usepackage{graphicx}
\usepackage{color}
\begin{document}

%\preprint{}

\title{Conformal Weyl gravity and perihelion precession}


\author{Joseph Sultana}
\email{joseph.sultana@um.edu.mt}
%\homepage[www.maths.um.edu.mt]
%\thanks{}
%\affiliation{Astrophysics Science Division, NASA/Goddard Space
%Flight Center, Greenbelt MD,20771 USA}
\affiliation{Department of Mathematics, University of Malta, Msida,
Malta}
%\email{joseph.sultana@um.edu.mt}
\author{Demosthenes Kazanas}
\email{demos.kazanas@nasa.gov} \affiliation{Astrophysics Science
Division, NASA/Goddard Space Flight Center, Greenbelt Maryland 20771
USA}
\author{Jackson Levi Said}
\email{jsai0004@um.edu.mt} \affiliation{Department of Physics,
University of Malta, Msida, Malta}

\date{\today}

\begin{abstract}

\baselineskip=14pt

We investigate the perihelion shift of planetary motion in conformal
Weyl gravity using the metric of the static, spherically symmetric
solution discovered by Mannheim \& Kazanas (1989). To this end we
employ a procedure similar to that used by Weinberg for the
Schwarzschild solution, which has also been used recently to study
the solar system effects of the cosmological constant $\Lambda$. We
show that besides the general relativistic terms obtained earlier
from the Schwarzschild - de Sitter solution, the expression for the
perihelion shift includes a negative contribution which arises from
the linear term $\gamma r$ in the metric. Using data for perihelion
shift observations we obtain constraints on the value of the
constant $\gamma$ similar to that obtained earlier using galactic
rotational curves.

\end{abstract}


\maketitle

\section{Introduction}

\baselineskip=15pt

One of the possible alternatives to the standard second order
Einstein's theory of gravity which has been proposed during the last
two decades is conformal Weyl gravity \cite{conformal1,
conformal2,conformal3}. Instead of fixing the gravitational action
{\textbf by} demanding that the theory be no higher than second order as in
the case of the Einstein-Hilbert action, Weyl gravity employs the
principle of local conformal invariance of spacetime as the
supplementary condition that fixes the gravitational action. This
means that the theory is invariant under local conformal stretching
of spacetime geometry of the form
\begin{equation}
g_{\mu\nu}(x) \rightarrow \Omega^2(x) g_{\mu\nu}(x),
\end{equation}
where $\Omega(x)$ is a smooth strictly positive function. This
restrictive conformal invariance leads to a fourth order theory with
a unique conformally invariant action
\begin{eqnarray} I_{W} & = & -\alpha\int
d^4x(-g)^{1/2}C_{\lambda\mu\nu\kappa}C^{\lambda\mu\nu\kappa} \nonumber\\
      & = & -2\alpha\int d^4x(-g)^{1/2}[R_{\mu\kappa}R^{\mu\kappa} -
      (R^{\nu}_{\nu})^2/3]+{\rm a~total~derivative}, \label{action}
\end{eqnarray}
where $C_{\lambda\mu\nu\kappa}$ is the conformal Weyl tensor and
$\alpha$ is a purely dimensionless coefficient. This action gives
rise \cite{conformal1} to the gravitational field equations given by
\begin{eqnarray}
\sqrt{-g}g_{\mu\alpha}g_{\nu\beta}\frac{\delta I_{W}}{\delta
g_{\alpha\beta}}
&=-2\alpha W_{\mu\nu}\nonumber\\
&=-\frac{1}{2}T_{\mu\nu}, \label{weyl_field_eqns}
\end{eqnarray}
where $T_{\mu\nu}$ is the stress-energy tensor, and
\begin{equation}
W_{\mu\nu} = 2C^{\alpha\ \ \beta}_{\ \mu\nu\ ;\alpha\beta} +
C^{\alpha\ \ \beta}_{\ \mu\nu}R_{\alpha\beta}, \label{w1}
\end{equation}
is the Bach tensor. When $R_{\mu\nu}$ is zero the tensor
$W_{\mu\nu}$ vanishes, so that any vacuum solution of Einstein's
field equations is also a vacuum solution of Weyl gravity. However
the converse is not necessarily true. Despite the highly nonlinear
character of the field equations, a number of exact solutions
\cite{furthersolutions1,furthersolutions2,furthersolutions3,
furthersolutions4} of conformal Weyl gravity have been found in the
case of spherical and axis symmetry. Recently cylindrically
symmetric solutions \cite{cylindrical1,cylindrical121} in Weyl gravity have also
been studied. Moreover by studying the interior structure of a
static spherically symmetric gravitational source, it was shown
\cite{furthersolutions4} that the field inside the source is
described exactly by a fourth order Poisson equation which still
admits a Newtonian potential $1/r$ term. Therefore although the
second order Poisson equation in General Relativity is sufficient to
generate a Newtonian potential, it is not by any means a necessary
requirement, so that Newton's law of gravity remains valid in the
fourth order Weyl gravity.

The exact static and spherically symmetric vacuum solution for
conformal gravity is given, up to a conformal factor, by the metric
\cite{conformal1}
\begin{equation}
ds^2 = -B(r)dt^2 + \frac{dr^2}{B(r)} + r^2(d\theta^2 + \sin^2\theta
d\phi^2), \label{generalmetric}
\end{equation}
where
\begin{equation}
B(r) = 1 - \frac{\beta(2 - 3\beta\gamma)}{r} - 3\beta\gamma + \gamma
r - k r^2, \label{eq:metric}
\end{equation}
and $\beta,\ \gamma,\  \mbox{and}\ k$ are integration constants.
This solution includes as special cases the Schwarzschild solution
$(\gamma = k = 0)$ and the Schwarzschild-de Sitter $(\gamma = 0)$
solution; the latter requiring the presence of a cosmological
constant in Einstein gravity. The constant $\gamma$ has dimensions
of acceleration, and so the solution provides a characteristic,
constant acceleration.  Although the magnitude of this integration
constant remains uncertain, it has been associated \cite{conformal1}
with the inverse Hubble length, i.e. $\gamma \simeq 1/R_H$.  In this
case the effects of the acceleration are comparable to those due to
the Newtonian potential term $2\beta/r \equiv r_s/r$ ($r_s$ is the
Schwarzschild radius), on length scales given by
\begin{equation}
r_s / r^2 \simeq \gamma \simeq 1/R_H ~~{\rm or}~~ r \simeq (r_s \,
R_H)^{1/2}. \label{eq:MRrelation}
\end{equation}
As noted in \cite{conformal1}, for a galaxy of mass $M \simeq
10^{11}\; {\rm M}_{\odot}$ with  $r_s \simeq 10^{16}$ cm and $R_H
\simeq 10^{28}$ cm, this scale is $r \sim 10^{22}$ cm, i.e. roughly
the size of the galaxy, a fact that prompted Mannheim \& Kazanas to
produce fits to the galactic rotation curves using the metric of Eq.
(\ref{eq:metric}) above. One should point out that Eq.
(\ref{eq:MRrelation}) describes not a particular length scale but a
continuum of sizes at which the contribution from the linear term
becomes significant. Objects along this sequence encompass not only
galaxies but, at larger scales also galaxy clusters and at lower
scales globular clusters, only recently found to require the
presence of dark matter in order to account for the observed
dynamics \cite{scarpa, kazanas08}. As we know, to account for the
dynamics of objects along this sequence within the standard
gravitational theory, one would need to invoke the presence of dark
matter.

The classical tests in Einstein's gravity for the metric
(\ref{eq:metric}) with vanishing $\gamma$ have been well documented
since the discovery of the theory, and therefore it would be natural
to study any effects of the linear term in the metric on such tests.
The issue of the propagation of null geodesics, particularly the
computation of the bending of light by a spherically symmetric
object using the metric in (\ref{eq:metric}) has been studied in
detail. The first results obtained in \cite{deflection1,
deflection2, deflection3, amore06} showed that besides the positive
Einstein deflection of $4\beta/b$, the expression for the deflection
of light in Weyl gravity, contained an extra term $-\gamma b$, where
$b$ is the impact parameter. This \textbf{led} to the paradoxical situation
where the bending angle in lensing increased with the light ray's
impact parameter with respect to the lens. This problem was later
solved in Ref. \cite{sultana}, where it was shown that when the
curvature of the background asymptotically non-flat geometry in
(\ref{eq:metric}) is taken into account the total bending angle is
given by
\begin{equation}
\Delta\psi = \frac{4\beta}{b} - \frac{2\beta^2\gamma}{b} -
\frac{kb^3}{2\beta}, \label{bendingformula}
\end{equation}
such that the contribution from the linear term in the metric is
inversely proportional to the impact parameter. Moreover its ratio
to that of the standard $1/r$ component is of order $\beta \gamma$,
which, given the associations and magnitudes of these constants for
a galaxy, we get $\beta \gamma \simeq 10^{-12}$, i.e. insignificant
for all practical purposes.

In this paper we study timelike geodesics in Weyl gravity,
particularly the effect of the linear term in the metric on the
perihelion shift. Then using the available data for planetary
perihelion shifts we get constraints on the magnitude of $\gamma$
similar to that obtained earlier \cite{conformal1} from the fitting
of galactic rotational curves. Earlier studies \cite{alpher, kerr,
cruz, hackmann} of the geodesic structure in the Schwarzschild - de
Sitter solution showed that the cosmological constant $\Lambda$
increases the perihelion shift in the Schwarzschild solution by
$\pi\Lambda a^3 (1 - \epsilon^2)^3/m$, where $a,\ \epsilon$ are the
length of the semi-major axis and eccentricity respectively, and $m
= GM/c^2;\ M$ being the mass of the source. Some authors
\cite{islam,cardona} have used this correction to the perihelion
shift of Mercury to obtain upper bounds for the cosmological
constant, while others \cite{thulsi,wright} showed that the effect
of the cosmological constant is only significant at large radii and
since it is unmeasurably small for Mercury's orbit it cannot be used
to limit the cosmological constant.

In section II we start with the differential equation representing
timelike geodesics and use a method similar to that used by Weinberg
for the Schwarzschild solution, to obtain an expression for the
perihelion shift of a test particle in the exterior geometry
described by the metric in (\ref{eq:metric}). Then in section III
the results are summarized and discussed.

\section{Timelike geodesics, perihelion precession}
The timelike orbits for the metric
\begin{equation}
ds^2 = -B(r) dt^2 + A(r) dr^2 + r^2(d\theta^2 + \sin^2\theta
d\phi^2),
\end{equation}
are given by \cite{weinberg}
\begin{equation}
\frac{A(r)}{r^4}\left(\frac{dr}{d\phi}\right)^2 + \frac{1}{r^2} -
\frac{1}{J^2 B(r)} = -\frac{E}{J^2}, \label{de}
\end{equation}
where $J$ and $E$ are constants of the motion. Since $\beta\gamma <<
1$ for simplicity we take
\begin{equation}
B(r) = A^{-1}(r) = 1 - \frac{2\beta}{r} + \gamma r - k r^2.
\end{equation}
Then the angular distance between the perihelion $r_{-}$ and
aphelion $r_{+}$ is given by
\begin{equation}
\phi(r_{+}) - \phi(r_{-}) =
\int_{r_{-}}^{r_{+}}A^{1/2}(r)\left[\frac{1}{J^{2}B(r)} -
\frac{E}{J^2} - \frac{1}{r^2}\right]^{-1/2}\frac{dr}{r^2}.
\label{distance}
\end{equation}
Using the fact that $dr/d\phi$ vanishes at $r_{-}$ and $r_{+}$, one
can derive the following values for the constants of the motion
\begin{equation}
E = \frac{\frac{r_{+}^2}{B(r_{+})} -
\frac{r_{-}^2}{B(r_{-})}}{r_{+}^2 - r_{-}^2},
\end{equation}
and
\begin{equation}
J^2 = \frac{\frac{1}{B(r_{+})} -
\frac{1}{B(r_{-})}}{\frac{1}{r_{+}^2} - \frac{1}{r_{-}^2}},
\end{equation}
such that the angular distance in (\ref{distance}) becomes
\begin{equation}
\hspace{-3mm}\phi(r_{+}) - \phi(r_{-}) =
\int_{r_{-}}^{r_{+}}A(r)^{1/2}\left[\frac{r_{-}^2(B^{-1}(r) -
B^{-1}(r_{-})) - r_{+}^2(B^{-1}(r) - B^{-1}(r_{+}))}{r_{+}^2 r_{-}^2
(B^{-1}(r_{+}) - B^{-1}(r_{-}))} -
\frac{1}{r^2}\right]^{-1/2}\frac{dr}{r^2}. \label{distance2}
\end{equation}
The perihelion precession per orbit is given by
\begin{equation}
\Delta\phi = 2|\phi(r_{+}) - \phi(r_{-})| - 2\pi.
\end{equation}
The expression in the square brackets in (\ref{distance2}) vanishes
at $r_{+}$ and $r_{-}$, and hence for slightly eccentric orbits we
can write \vspace{2mm}
\begin{equation}
\left[\frac{r_{-}^2(B^{-1}(r) - B^{-1}(r_{-})) - r_{+}^2(B^{-1}(r) -
B^{-1}(r_{+}))}{r_{+}^2 r_{-}^2 (B^{-1}(r_{+}) - B^{-1}(r_{-}))} -
\frac{1}{r^2}\right] \approx C\left(\frac{1}{r_{-}} -
\frac{1}{r}\right)\left(\frac{1}{r} - \frac{1}{r_{+}}\right),
\end{equation}
where $C$ is a constant. Then letting $u = 1/r$ and differentiating
twice with respect to $u$ gives
\begin{equation}
C \approx 1 - \frac{(u_{+} - u_{-})(u_{+} + u_{-})A''(u)}{2(A(u_{+})
- A(u_{-}))}\Big|_{u=L^{-1}},
\end{equation}
where $L = 2/(u_{+} + u_{-}) = a(1 - \epsilon^2)$ is the
\textit{semilatus rectum} of the elliptic orbit. Again for slightly
elliptic orbits we can write
\begin{equation}
A(u_{+}) - A(u_{-}) \approx (u_{+} - u_{-}) A'\left(\frac{u_{+} +
u_{-}}{2}\right),
\end{equation}
so that
\begin{equation}
C \approx 1 - \frac{u A''(u)}{A'(u)}\Big|_{u = L^{-1}}.
\label{constant}
\end{equation}
Now
\begin{equation}
A(u) = B^{-1}(u) = 1 + 2\beta u + 4\beta^2 u^2 + \frac{4\beta k}{u}
- \frac{\gamma}{u} + \frac{k}{u^2} - 4\beta\gamma.
\end{equation}
Hence substituting this in (\ref{constant}) we get
\begin{equation}
C \approx \frac{2\beta + 3\gamma u^{-2} - 12 k\beta u^{-2} - 8k
u^{-3}}{2\beta + 8\beta^2 u  - 4 k\beta u^{-2} + \gamma u^{-2} - 2k
u^{-3}}\Big|_{u = L^{-1}},
\end{equation}
or
\begin{equation}
C \approx 1 - 4\beta u - \frac{2\gamma}{u} + \frac{\gamma}{\beta
u^2} + \frac{4k}{u^2} - \frac{3k}{\beta u^3}\Big|_{u = L^{-1}}.
\end{equation}
The expression for the angular distance in (\ref{distance2}) can be
written as
\begin{equation}
\phi(r_{+}) - \phi(r_{-}) =
-\int_{u_{-}}^{u_{+}}\frac{A(u)^{1/2}\,du}{[C(u_{-} - u)(u -
u_{+})]^{1/2}}.
\end{equation}
Using the substitution
\begin{equation}
u = \frac{1}{2}(u_{+} + u_{-}) + \frac{1}{2}(u_{+} - u_{-})\sin\psi,
\end{equation}
simplifies the integral to
\begin{equation}
\phi(r_{+}) - \phi(r_{-}) = \frac{1}{C^{1/2}}\int_{-\pi/2}^{\pi/2}
A(\psi)^{1/2}\,d\psi.
\end{equation}
This leads to
\begin{eqnarray}
\phi(r_{+}) - \phi(r_{-}) & \approx & \pi(1 + \frac{3\beta}{a(1 -
\epsilon^2)} + \frac{17}{2}ka^2(1 - \epsilon^2)^2 +
\frac{3k}{2\beta}a^3(1 - \epsilon^2)^3 \nonumber \\ & &
-\frac{\gamma}{2\beta}a^2(1 - \epsilon^2)^2 - 2\gamma a(1 -
\epsilon^2)),
\end{eqnarray}
so that the perihelion shift per orbit is given by
\begin{equation}
\Delta\phi \approx \frac{6\pi\beta}{a(1 - \epsilon^2)} +
\frac{3\pi}{\beta}ka^3(1 - \epsilon^2)^3 - \frac{\pi}{\beta}\gamma
a^2 (1 - \epsilon^2)^2. \label{shift}
\end{equation}

\section{Discussion and Conclusion}
The expression for the perihelion precession in conformal Weyl
gravity obtained in (\ref{shift}) constitutes the main result of
this paper. Besides the conventional term $6\pi\beta/a(1 -
\epsilon^2)$ representing the precession in the Schwarzschild
geometry, the expression includes two other terms that originate due
to the large scale structure of the embedding space time. The
$k$-dependent term is clearly associated with the $kr^2$ de Sitter
term of the metric of Weyl gravity, and is the same contribution
arising from the cosmological constant obtained earlier in
\cite{alpher, kerr, cruz, hackmann} for the Schwarzschild - de
Sitter solution. This contribution was expected considering the fact
that the metric in (\ref{eq:metric}) reduces to the Schwarzschild-de
Sitter solution when $\gamma$ approaches zero. The additional term
$\pi\gamma a^2 (1 - \epsilon^2)^2/\beta$ in (\ref{shift}) originates
from the linear $\gamma r$ term in the metric, and due to its
negative sign it reduces the amount of precession per orbit. This
has also been observed before in \cite{sultana} (see Eq.
(\ref{bendingformula}) in this paper), where it was shown that both
the linear and de Sitter terms in the metric diminish the light
bending angle. In this case however, the effects on the precession
from these two terms are opposite.

The nature and magnitude of the constant $\gamma$ are still unknown.
When $\beta = 0$ or when $r$ is sufficiently large so that all
$\beta$ dependent terms in (\ref{eq:metric}) can be ignored, the
metric can be re-written under the coordinate transformation
\begin{equation}
\rho = \frac{4r}{2(1 + \gamma r - k r^2)^{1/2} + 2 + \gamma r},
\quad \mbox{and} \quad \tau = \int R(t)dt,
\end{equation}
in the form
\begin{equation}
\hspace{-3mm} ds^2 = \frac{[1 - \rho^2(\gamma^2/16 +
k/4)]^2}{R^2(\tau)[(1 - \gamma\rho/4)^2 +
k\rho^2/4]^2}\left[-d\tau^2 + \frac{R^2(\tau)}{[1 -
\rho^2(\gamma^2/16 + k/4)]^2}(d\rho^2 + \rho^2(d\theta^2 +
\sin^2\theta d\phi^2))\right].
\end{equation}
The metric is therefore asymptotically conformal to a FLRW metric
with arbitrary scale factor $R(\tau)$ and spatial curvature $\kappa
= - k - \gamma^2/4$, i.e., it describes a spherically symmetric
object embedded in a conformally flat background space. The fact the
curvature of this background space depends on $\gamma$ and $k$
points towards a cosmological origin of $\gamma$. This \textbf{leads} to its
association with the inverse Hubble length in Ref.
\cite{conformal1}, so that $\gamma \sim 10^{-28}\  \mbox{cm}^{-1}$,
and this explained quite successfully the flat rotation curves in
galaxies and galaxy clusters.

On the other hand there is nothing in the theory which denies
$\gamma$ from being system dependent (i.e. like the mass $\beta$),
and in this case one can suggest that the linear term $\gamma r$ in
the metric provides the necessary changes in the spacetime geometry
to allow the embedding of a spherically symmetric matter
distribution in a cosmological background. Assuming a spatially flat
matter dominated universe with $k = 0$, one can use (\ref{shift}) to
get an upper bound for the magnitude of $\gamma$ from the difference
between the observed and the general relativistic values of the
precession of perihelia. So for example in the case of Mercury
\cite{pitjeva}
\begin{equation}
\Delta\phi_{\mbox{obs}} - \Delta\phi_{\mbox{gr}} = -0.0036 \pm
0.0050\ \mbox{arcsec/century}, \label{mercury}
\end{equation}
where $\Delta\phi_{\mbox{obs}}$ refers to the observed value of the
perihelion precession once \textbf{corrected} for the general precession of
the equinoxes and for the perturbations due to other planets. The
contribution to the precession of the perihelion from the linear
term in the metric can be written in the form
\begin{equation}
\Delta\phi_{\gamma} = -\frac{2\pi\gamma L^2}{r_{s}},\label{residual}
\end{equation}
where $r_{s} = 2\beta = \frac{2GM}{c^2}$. For Mercury $r_{s}L^{-1} =
5.3\times10^{-8}$ so that
\begin{equation}
\Delta\phi_{\gamma} = 1.2\times10^8\gamma L\ \mbox{rad/orbit}.
\end{equation}
Therefore from (\ref{mercury}) we get $\gamma < 1.5\times10^{-31}\
\mbox{cm}^{-1}$, which is about three orders of magnitude smaller
than the large scale estimate obtained from the fitting of galactic
rotational curves. A similar calculation for other planets such as
the Earth leads to tighter upper bounds for $\gamma$. However when
comparing bounds on $\gamma$  derived from planetary data with those
obtained from the large scale geometry of the universe, one should
keep in mind that the calculation of the perihelion precession
assumes that planets are treated as test particles devoid of
self-gravity, and so this weakens the validity of such bounds. In
fact in the case of the asteroid Icarus with a diameter of just $1
\mbox{km}$, (so that the effect of self-gravity in this case is
insignificant) the difference between the observed and general
relativistic perihelion precession is given by \cite{weinberg}
\begin{equation}
\Delta\phi_{\mbox{obs}} - \Delta\phi_{\mbox{gr}} = -0.5 \pm 0.8\
\mbox{arcsec/century}, \label{icarus}
\end{equation}
which gives using (\ref{residual}), the value $\gamma <
1.3\times10^{-28}\ \mbox{cm}^{-1}$. This fits nicely with the
cosmological estimate of $10^{-28}\ \mbox{cm}^{-1}$ obtained earlier
from galactic rotational curves.

\begin{acknowledgments}
J.S. gratefully acknowledges financial support from the University
of Malta during his visit at NASA-GSFC.
\end{acknowledgments}

\end{document}